\documentclass[aps,twocolumn,showpacs,superscriptaddress,nofootinbib,groupedaddress]{revtex4} 
\usepackage{pstricks}
\usepackage{color}
\usepackage{graphicx}
\usepackage{amsmath}
\usepackage{amssymb}
\usepackage[colorlinks=true,allcolors=darkred, pdfborder={0 0 0}]{hyperref}
\usepackage[T1]{fontenc}
\usepackage[latin9]{inputenc}
\usepackage{array}
\usepackage{booktabs}
\usepackage{mathrsfs}
\usepackage[normalem]{ulem}

\definecolor{darkred}{rgb}{0.6,0,0}
\newcommand{\changed}[1]{\textcolor{darkred}{#1}}

\definecolor{darkred}{rgb}{0.6471, 0.1098, 0.1882}

\def\gsim{\raise0.3ex\hbox{$\;>$\kern-0.75em\raise-1.1ex\hbox{$\sim\;$}}}
\def\lsim{\raise0.3ex\hbox{$\;<$\kern-0.75em\raise-1.1ex\hbox{$\sim\;$}}}

\begin{document}

\title{Are the B decay anomalies related to neutrino oscillations?}
\author{Sofiane M. Boucenna}
\email{boucenna@lnf.infn.it}
\address{INFN, Laboratori Nazionali di Frascati, C.P. 13, 100044 Frascati, Italy.}
\author{Jos\'e W.F. Valle}
\email{valle@ific.uv.es}
\address{Instituto de F\'{\i}sica Corpuscular (CSIC-Universitat de Val\`{e}ncia), Apdo. 22085, E-46071 Valencia, Spain.}
\author{Avelino Vicente}
\email{avelino.vicente@ulg.ac.be}
\address{Instituto de F\'{\i}sica Corpuscular (CSIC-Universitat de Val\`{e}ncia), Apdo. 22085, E-46071 Valencia, Spain.}
\address{IFPA, Dep. AGO, Universit\'e Li\`ege, Bat B5, Sart-Tilman B-4000, Li\`ege 1, Belgium}
%



\begin{abstract}
\noindent

Neutrino oscillations are solidly established, with a hint of CP
violation just emerging. Similarly, there are hints of lepton
universality violation in $b \to s$ transitions at the level of $2.6
\sigma$.
By assuming that the unitary transformation between weak and mass
charged leptons equals the leptonic mixing matrix measured in neutrino
oscillation experiments, we predict several lepton flavor violating
(LFV) B meson decays. We are led to the tantalizing possibility that
some LFV branching ratios for B decays correlate with the leptonic CP
phase $\delta$ characterizing neutrino oscillations.
Moreover, we also consider implications for $\ell_i \to
\ell_j \ell_k \ell_k$ decays.

\end{abstract}

\maketitle

\section*{Introduction}

The historical discovery of the Higgs
boson~\cite{Aad:2012tfa,Chatrchyan:2012ufa} would have completed our
picture of particle physics, were it not for the solid evidence we now
have that neutrino flavors interconvert~\cite{Maltoni:2004ei}. Apart
from neutrino oscillations and cosmology, no other signs of new
physics (NP) have been established. However, some indirect signs might
have been found by the LHCb collaboration. In 2013, they have
published the results of the measurement of a variety of observables
in $b \to s$ transitions. In some cases, the experimental result was
found to be in clear tension with the Standard Model (SM)
prediction. These include angular observables \changed{\cite{Matias:2012xw,DescotesGenon:2012zf,Descotes-Genon:2013vna,Descotes-Genon:2013wba}} in $B
\to K^\ast \mu^+ \mu^-$~\cite{Aaij:2013qta}, as well as a sizable
suppression of several branching
ratios~\cite{Aaij:2014pli,Aaij:2013aln}. Recently, the LHCb announced
new results based on the complete LHC Run I
dataset~\cite{LHCbtalk}. The inclusion of new data has confirmed the
anomalies, which are currently at the $\sim 4 \sigma$
level. Furthermore, in 2014, the LHCb collaboration found an
intriguing indication of lepton universality violation in the
ratio~\cite{Aaij:2014ora}
\begin{equation}
R_K = \frac{\text{BR}(B \to K \mu^+ \mu^-)}{\text{BR}(B \to K e^+ e^-)} 
= 0.745^{+0.090}_{-0.074} \pm 0.036 \, .
\end{equation}
This experimental measurement, obtained in the low dilepton invariant
mass regime, is $2.6 \sigma$ away from the SM result $R_K^{SM} =
1.0003 \pm 0.0001$~\cite{Hiller:2003js}. Although the statistical
significance of this discrepancy is not enough to claim a discovery,
it is highly suggestive that several independent global fits
\cite{Alonso:2014csa,Hiller:2014yaa,Ghosh:2014awa,Hurth:2014vma,Altmannshofer:2014rta}
have shown that this hint can be explained by the same type of new
physics contributions as the previous $b \to s$ anomalies.

The violation of lepton universality usually comes together with the
violation of lepton flavor. Based on symmetry arguments, Glashow,
Guadagnoli and Lane~\cite{Glashow:2014iga} recently argued that the
observation of universality violation in the lepton flavor conserving
(LFC) $B \to K \ell_i^+ \ell_i^-$ decays implies the existence of the
lepton flavor violating (LFV) processes $B \to K \ell_i^+ \ell_j^-$
(with $i \neq j$). The idea of LFV in $B$ meson decays has been
further explored
in~\cite{Gripaios:2014tna,Bhattacharya:2014wla,Sahoo:2015wya,Varzielas:2015iva,Alonso:2015sja}.

Here we take a step further in this direction. Since we lack a theory
a flavor, we can not make definite predictions for the LFV rates using
the LFC ones as input. Hence we make the simplest alternative
assumption, namely, that the unitary transformation between weak and
mass charged lepton states is given by the leptonic mixing matrix
measured in neutrino oscillation experiments. Under this assumption we
make numerical predictions for several LFV observables in the B
system. We emphasize that this assumption is not completely {\it ad
  hoc}. It will actually be a prediction in models where the leptonic
mixing arises from the charged lepton sector. We refer to
\cite{Altarelli:2004jb} for a general discussion and an example model.

\section*{General aspects of the $b \to s$ anomalies}

The effective hamiltonian describing $b \to s$ transitions can be
expressed as:
\begin{equation}
\mathcal H_{\text{eff}} = - \frac{4 G_F}{\sqrt{2}} \, V_{tb} V_{ts}^\ast \, \frac{e^2}{16 \pi^2} \, \sum_i \left(C_i \mathcal O_i + C^\prime_i \mathcal O^\prime_i \right) + \text{h.c.}
\end{equation}
Here $G_F$ is the Fermi constant, $e$ the electric charge and $V$ the
CKM matrix. The Wilson coefficients $C_i$ and $C^\prime_i$ encode the
different (SM and NP) contributions to the effective operators
$\mathcal O_i$ and $\mathcal O^\prime_i$. The analysis of the
available experimental data on $b \to s$ transitions reveals that the
effective operators relevant for the resolution of the $b \to s$
anomalies are:
\begin{align}
\mathcal O_9 \equiv \mathcal O_9^{\mu \mu} &= \left( \bar s \gamma_\alpha P_L b \right) \, \left( \bar \mu \gamma^\alpha \mu \right) \, , \label{eq:O9} \\
\mathcal O_{10} \equiv \mathcal O_{10}^{\mu \mu} &= \left( \bar s \gamma_\alpha P_L b \right) \, \left( \bar \mu \gamma^\alpha \gamma_5 \mu \right) \,, \label{eq:O10}
\end{align}
where $P_L = \frac{1}{2} (1-\gamma_5)$ is the left-chirality
projector. Several independent global fits
\cite{Hiller:2014yaa,Ghosh:2014awa,Hurth:2014vma,Altmannshofer:2014rta}
find a significant tension between the SM results for the Wilson
coefficients of these operators and the experimental data. This can be
clearly alleviated in the presence of NP contributions. According to
the global fit \cite{Altmannshofer:2014rta}, the $C_9^{\mu
  \mu}$ coefficient is the key to improve the fits. More precisely,
one finds a reasonable agreement with data when NP provides a negative
contribution to $\mathcal O_9^{\mu \mu}$, with $C_9^{\mu \mu,
  \text{NP}} \sim - 30 \% \times C_9^{\mu \mu, \text{SM}}$. Similar
improvements are found when NP enters in the $SU(2)_L$ invariant
direction $C_9^{\mu \mu, \text{NP}} = - C_{10}^{\mu \mu, \text{NP}}$,
with $C_9^{\mu \mu , \text{NP}} \sim - 12 \% \times C_9^{\mu \mu, \text{SM}}$.

\section*{Predicting lepton flavor violation in $B$ meson decays}
\label{sec:leptonmixing}

Here we raise the following question: can the leptonic mixing matrix
provide the required lepton flavor structure in $\mathcal O_9$ and
$\mathcal O_{10}$? And if so, what are the predictions for lepton
flavor violation in the $B$ sector?
As suggested by global fits, let us assume that the relevant NP
operator contains a left-handed leptonic current. In this case, this
operator can be generally written in the mass basis as:
\begin{equation} \label{eq05}
\mathcal O^{ij} = \frac{1}{\Lambda^2} J_\alpha^d J_{\ell_{ij}}^\alpha \, , 
\end{equation}
where
\begin{eqnarray}
J_\alpha^d &=& C^Q_{bs} \, \bar b \gamma_\alpha P_L s \, , \label{eq:JQ} \\
J_{\ell_{ij}}^\alpha &=& C^L_{ij} \, \bar \ell_i \gamma^\alpha P_L \ell_j \, , \label{eq:JL}
\end{eqnarray}
and $\Lambda$ is the energy scale of the NP inducing this
operator. The $i,j$ indices denote the lepton flavor combination
characterizing the operator in eq. \eqref{eq:JL}. The $3 \times 3$
matrices $C^Q$ and $C^L$ completely determine the relations among the
Wilson coefficients for different flavor choices. On the other hand,
in the interaction (gauge) basis, $\mathcal O$ takes the same form,
but the quark and lepton currents are written in terms of gauge
eigenstates $d^\prime$ and $\ell^\prime$ as
\begin{eqnarray}
J_\alpha^d &=& \tilde C^Q_{mn} \, \bar d_m^\prime \gamma_\alpha P_L d_n^\prime \, , \label{eq:JQgauge} \\
J_{\ell_{ij}}^\alpha &=& \tilde C^L_{ij} \, \bar \ell_i^\prime \gamma^\alpha P_L \ell_j^\prime \, . \label{eq:JLgauge}
\end{eqnarray}

We now focus on the leptons. By combining eqs. \eqref{eq:JL} and
\eqref{eq:JLgauge} one finds the relation between $C^L$ and $\tilde
C^L$,
\begin{equation} \label{eq:Cs}
C^L = U_\ell^\dagger \tilde C^L U_\ell \, ,
\end{equation}
where $U_\ell$ is the unitary matrix which relates the left-handed
charged lepton gauge and mass eigenstates as $\ell^\prime = U_\ell
\ell$. Similarly, the left-handed neutrino gauge and mass eigenstates
are connected by another matrix, $U_\nu$, as $\nu^\prime = U_\nu
\nu$. The product of these two matrices determines the leptonic
charged current weak interaction,
\begin{eqnarray}
\mathcal L_{\text{cc}} &=& - \frac{g}{2 \sqrt{2}} \left[ W_\mu^- \, \bar \ell^\prime \gamma^\mu P_L \nu^\prime + \text{h.c.} \right] \nonumber \\
&=& - \frac{g}{2 \sqrt{2}} \left[ W_\mu^- \, \bar \ell \gamma^\mu K P_L \nu + \text{h.c.} \right] \, ,
\end{eqnarray}
where $K = U_\ell^\dagger U_\nu$ is the leptonic mixing matrix
measured in neutrino oscillation experiments. If $U_\nu = \mathbb{I}$,
the left-handed neutrino gauge and mass eigenstates are the same and
all the mixing is in the left-handed charged leptons. In this case $K
= U_\ell^\dagger$ and eq. \eqref{eq:Cs} leads to
\begin{equation} \label{eq:Cs2}
C^L = K \tilde C^L K^\dagger \, .
\end{equation}

We do not attempt to give any model prediction for $\tilde
C^L$. Instead, we will assume that it is diagonal but with
non-universal entries. In that case one can determine the required
$\tilde C^L$ which, after using eq. \eqref{eq:Cs2}, leads to a $C^L$
matrix compatible with the observations in $b \to s$ transitions. In
particular, the resulting $C^L$ must have a strong hierarchy between
the $ee$ and $\mu \mu$ entries, $C^L_{ee} \ll C^L_{\mu \mu}$, in order
to induce a sizable correction to $B \to K^{(*)} \mu^+ \mu^-$ and a
negligible one to $B \to K^{(*)} e^+ e^-$. \\

\begin{center}
\textit{``Deriving'' $C^L$ from neutrino oscillations}
\end{center}

Barring tuning of the parameters, we find two generic $\tilde C^L$ matrices in the
gauge basis that lead to valid $C^L$ matrices in the mass basis. Their
forms define our two scenarios:
\begin{itemize}
\item {\bf Scenario A}: $\tilde C^L = \text{diag} (0,\epsilon,1)$
\item {\bf Scenario B}: $\tilde C^L = \text{diag} (\epsilon,0,1)$
\end{itemize}
Here $\epsilon \ll 1$ is a small parameter~(interestingly enough, note that
Ref.~\cite{Glashow:2014iga} considered $\tilde C^L = \text{diag}
(0,0,1)$, which corresponds to any of our scenarios in the limit $\epsilon =
0$)~\footnote{We note that a mixture of our two scenarios, i.e., $\tilde
  C^L=\mathrm{diag}(\epsilon,\epsilon,1)$, with $\epsilon \ll 1$, is also viable.}. Using the standard parameterization for
the leptonic mixing matrix $K$, one finds that in order to suppress
the contributions to the $ee$ Wilson coefficients, $\epsilon$ must be
close to
\begin{align}
\epsilon_{\text{A}} =& - \frac{\tan^2 \theta_{13}}{\sin^2 \theta_{12}} \quad & \text{in scenario A} \, , \label{eq:epsilonbestA} \\
\epsilon_{\text{B}} =& - \frac{\tan^2 \theta_{13}}{\cos^2 \theta_{12}} \quad & \text{in scenario B} \, . \label{eq:epsilonbestB}
\end{align}
Taking $3 \sigma$ ranges for the mixing angles from the latest global
fit to neutrino oscillation data~\cite{Forero:2014bxa}, one finds the
ranges $\left[ -0.10 , -0.05 \right]$ for scenario A and $\left[ -0.05
  , -0.03 \right]$ for scenario B, irrespective of the neutrino mass
spectrum; normal and inverted hierarchies giving basically the same
results. Interestingly, $\theta_{13} \neq 0$ implies $\epsilon \neq
0$, indicating a suggestive connection between quarks and leptons. 

We can now obtain $C^L$ for both scenarios. Let us first consider case
A.  Assuming $\epsilon = \epsilon_{\text{A}}$ and taking the best-fit
values from \cite{Forero:2014bxa}, we find:
\begin{widetext}
\begin{equation}
C^L = \left(
\begin{array}{ccc}
 0 & -0.023+0.117 \, e^{i \delta} & 0.026 +0.102 \, e^{i \delta} \\
 -0.023+0.117 \, e^{-i \delta} & 0.005 \cos \delta+0.532 & -0.001 \cos \delta + 0.005 \, i \, \sin \delta +0.509 \\
 0.026 +0.102 \, e^{-i \delta} & -0.001 \cos \delta - 0.005 \, i \, \sin \delta +0.509 & 0.394 -0.005 \cos \delta
\end{array}
\right) \, ,
\end{equation}
\end{widetext}
where $\delta$ is the Dirac leptonic CP violating phase.  In the CP
conserving case ($\delta = 0$) this matrix simplifies to
\begin{equation}
C^L = \left(
\begin{array}{ccc}
 0 & 0.094 & 0.128 \\
 0.094 & 0.537 & 0.508 \\
 0.128 & 0.508 & 0.389
\end{array}
\right) \, .
\end{equation}
{Regarding case B, assuming now $\epsilon = \epsilon_{\text{B}}$ and
  taking the best-fit values for the mixing angles from
  \cite{Forero:2014bxa}, one finds
\begin{widetext}
\begin{equation}
C^L = \left(
\begin{array}{ccc}
 0 & 0.011 +0.117 \, e^{i \delta} & -0.012+0.102 \, e^{i \delta} \\
 0.011 +0.117 \, e^{-i \delta} & 0.548 - 0.003 \cos \delta & - 0.003 \, i \, \sin  \delta +0.489 \\
 -0.012+0.102 \, e^{-i \delta} & 0.003 \, i \, \sin \delta + 0.489 & 0.003 \cos \delta +0.416
\end{array}
\right) \, .
\end{equation}
\end{widetext}
In the CP conserving case ($\delta = 0$) this matrix
simplifies to
\begin{equation}
C^L = \left(
\begin{array}{ccc}
 0 & 0.128 & 0.090 \\
 0.128 & 0.545 & 0.489 \\
 0.090 & 0.489 & 0.419
\end{array}
\right) \, .
\end{equation}

\begin{figure}[t!]
\centering \includegraphics[width=0.5\textwidth]{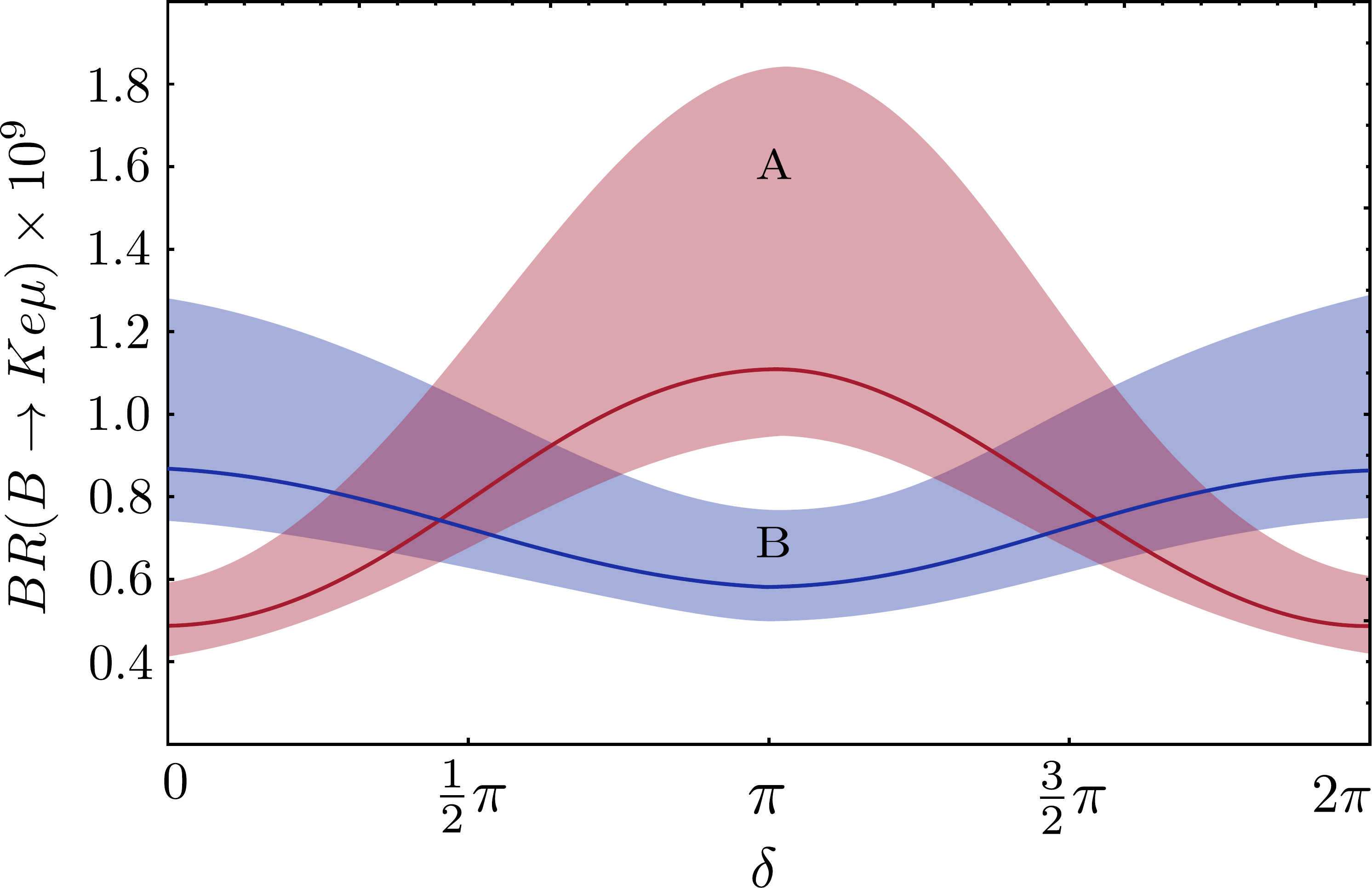}
\caption{The branching ratio of the decay $B \to K e \mu$ versus the
  CP violating phase $\delta$ in scenarios A and B. The bands are
  obtained by taking the leptonic mixing angles within their $1 \sigma
  $ range w.r.t. the best-fit value (solid
  line)~\cite{Forero:2014bxa}.}
\label{fig:BKemu-A}
\end{figure}

Comparing the $C^L$ matrices for our two scenarios, we find that they are of the same of order of magnitude and the most significant difference lies in the terms involving $\delta$. This is what will allow us to relate $B$ decays to the leptonic CP phase. }

\section*{ Lepton flavor violation in the B system}

The matrix $C^L$ can be used to make definite predictions for ratios
of branching ratios in $B \to K \ell_i^+ \ell_j^-$ decays,
\begin{equation} \label{eq:BRsB}
\text{BR}(B \to K \ell_i^\pm \ell_j^\mp) = 2 \, \rho_{\text{NP}}^2 \, \Phi_{ij} \, \left| \frac{C^L_{ij}}{C^L_{\mu \mu}} \right|^2 \, \text{BR}(B \to K \mu^+ \mu^-) \, .
\end{equation}
Here $\text{BR}(B \to K \ell_i^\pm \ell_j^\mp) = \text{BR}(B \to K
\ell_i^+ \ell_j^-) + \text{BR}(B \to K \ell_i^- \ell_j^+)$ and
$\text{BR}(B \to K \mu^+ \mu^-) = (4.29 \pm 0.22) \times 10^{-7}$ is
the LHCb result~\cite{Aaij:2014pli}, measured using the $3$ fb$^{-1}$
dataset after LHC Run I in the complete $q^2$ range, where $q^2 =
M_{\mu \mu}^2$ is the dimuon invariant mass. The factor
$\rho_{\text{NP}}$ is the NP fraction of the $B \to K \mu^+ \mu^-$
amplitude, $\rho_{\text{NP}} = \mathcal M_{\text{NP}} / \mathcal
M_{\text{Total}}$~\cite{Glashow:2014iga}. Using the results of the
global fit \cite{Altmannshofer:2014rta}, which gives $C_9^{\mu \mu ,
  \text{NP}} \sim - 12 \% \times C_9^{\mu \mu, \text{SM}}$,
$\rho_{\text{NP}}$ is found to be $\rho_{\text{NP}} \sim -
0.136$~\footnote{The authors of \cite{Glashow:2014iga} derive their
  value for $\rho_{\text{NP}}$ from the LHCb $R_K$ measurement,
  obtaining $\rho_{\text{NP}} \sim - 0.159$.}. Finally, the
$\Phi_{ij}$ factor accounts for phase space {and charged lepton mass}
effects. These introduce sizable corrections for final states
including $\tau$ leptons. Using the results of
Ref.~\cite{Crivellin:2015era}, we find $\Phi_{\mu e} \simeq 1$ and
$\Phi_{\tau e} = \Phi_{\tau \mu} \simeq 0.63$. Finally, we note that
the parameterization in terms of $\rho_{\text{NP}}$ is only exact in
the limit of vanishing non-factorizable contributions. However, we
have found that these corrections are negligible for the processes we
are interested in. \\

For $\delta = 0$, we obtain the following predictions for the $B \to
K$ LFV transitions in scenario A,
\begin{eqnarray}
\text{BR}(B \to K e^\pm \mu^\mp) &\in& \left[ 4.3 , 6.2 \right] \times 10^{-10} \, , \\
\text{BR}(B \to K e^\pm \tau^\mp) &\in& \left[ 0.4 , 1.3 \right]  \times 10^{-9} \, , \\
\text{BR}(B \to K \mu^\pm \tau^\mp) &\in& \left[ 0.8 , 1.6 \right]  \times 10^{-8} \, .
\end{eqnarray}
These have been derived using the LHCb central value and taking the
leptonic mixing angles in the preferred $1 \sigma$ ranges found by the
fit \cite{Forero:2014bxa}. The main generic prediction from our setup
is thus
\begin{equation}
\text{BR}(B \to K \mu^\pm \tau^\mp) \gg \text{BR}(B \to K e^\pm \mu^\mp) , \text{BR}(B \to K e^\pm \tau^\mp) \, .
\end{equation}

{ However, experimentally the decay $B \to K e^\pm \mu^\mp$ is the
  easiest to search for and reconstruct.  Indeed, electron and tau
  final states are, $\mathcal{O}(20\%)$ and $\mathcal{O}(80\%)$
  respectively, worse to reconstruct.  Moreover future RUN II data
  will probe the region of $\mathcal{O}(10^{-10})$ for this channel
  providing a test of our scenario.  }

\begin{figure}[t!]
\centering \includegraphics[width=0.5\textwidth]{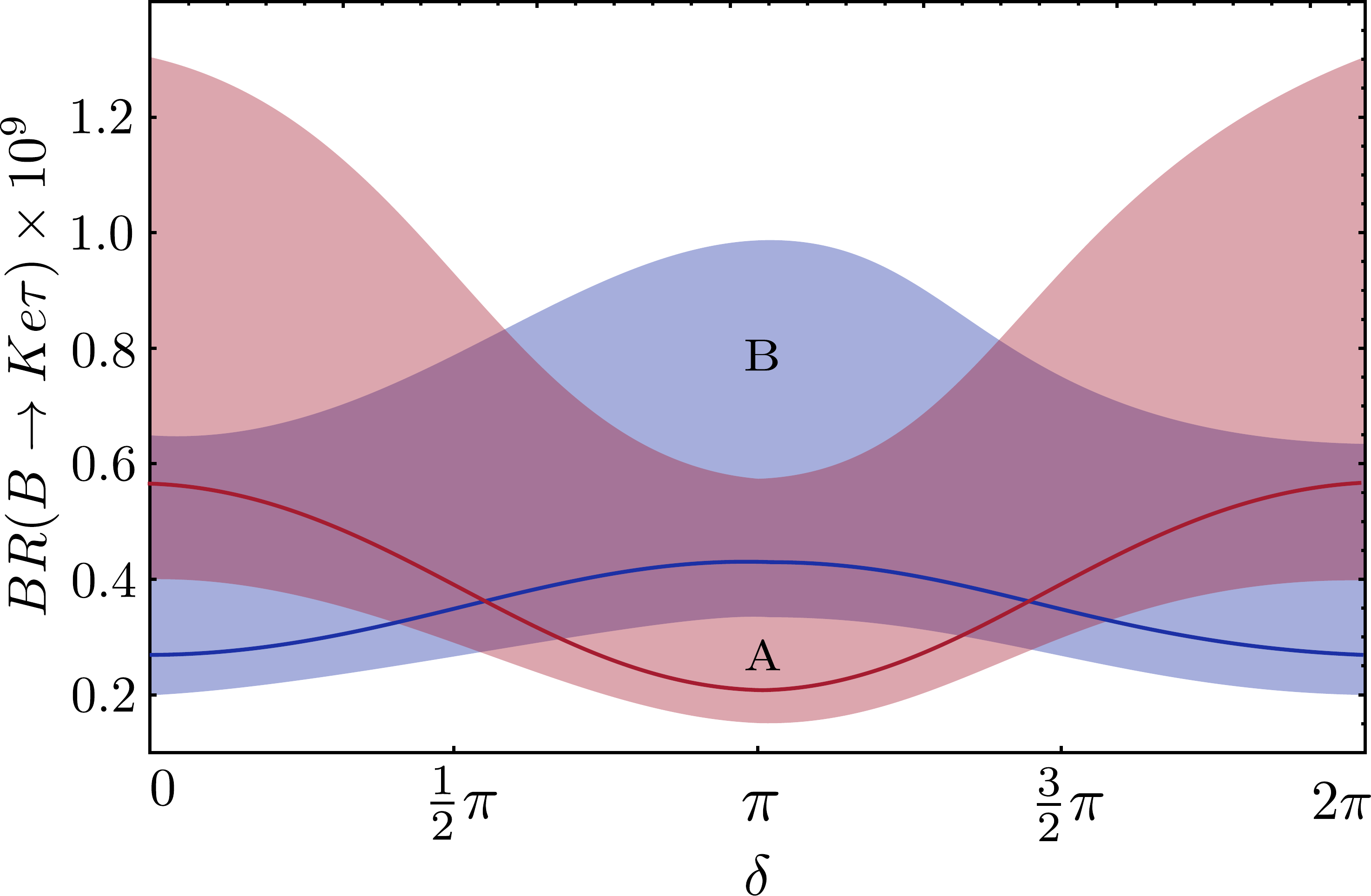}
\caption{Same as fig.~(\ref{fig:BKemu-A}) for the branching ratio of
  the decay $B \to K e \tau$.}
\label{fig:BKetau-A}
\end{figure}

\section*{Rare B decays and leptonic CP violation}

One can now consider a scenario with a non-zero value of the CP
violating phase $\delta$ characterizing neutrino oscillations. In this
case, we are led to the fascinating possibility that the LFV branching
ratios for B meson decays will depend upon $\delta$. Our results can
be found in figs.~(\ref{fig:BKemu-A}) and (\ref{fig:BKetau-A})
corresponding to the decay modes $B \to K \mu^\pm e^\mp$ and $B \to K
\tau^\pm e^\mp$ respectively.  This would suggest an alternative way of
probing $\delta$ by using LFV $B$ meson decays.

\section*{$\boldsymbol{\ell_i \to \ell_j \ell_k \ell_k}$ decays}

The same strategy can be extended to other LFV observables if induced
mainly by vectorial operators, as in eqs. \eqref{eq:O9} and
\eqref{eq:O10}. Assuming the same leptonic currents, the analogous
operators for the purely leptonic LFV processes $\ell_i \to
\ell_j \ell_k \ell_k$ are:
\begin{equation}
\mathcal O_{4 \ell} = \frac{1}{\Lambda^2} \left( C^L_{ij} \, \bar \ell_i \gamma_\alpha P_L \ell_j \right) \, \left( C^L_{mn} \, \bar \ell_m \gamma^\alpha P_L \ell_n \right) \, .
\end{equation}
Here we assume that the scale of the NP responsible for the vectorial
LFV operators is the same as the one relevant for $B$ meson decays,
eq.~(\ref{eq05}), although in full generality these could be
unrelated.
The flavor structure of $\mathcal O_{4 \ell} \equiv \mathcal O_{4
  \ell}^{ijmn}$ is given by the product $C^L_{ij} C^L_{mn}$ which,
following the same prescription as for the $B$ meson decays, can be
written as $C^L_{ij} C^L_{mn} = \left( K \tilde C^L K^\dagger
\right)_{ij} \left( K \tilde C^L K^\dagger \right)_{mn}$.

The $\mathcal O_{4 \ell}$ operator induces several $\ell_i \to \ell_j
\ell_k \ell_k$ decay processes: (i) $\ell^-_i \to \ell^-_j \ell^-_k
\ell^+_k$, 
and (ii) $\ell^-_i \to \ell^+_j \ell^-_k \ell^-_k$ (with $k \neq j$).
Their branching ratios can be written as~\cite{Abada:2014kba}
\begin{equation}
\text{BR}(\ell_i \to \ell_j \ell_k \ell_k) = \kappa \, \frac{m_{\ell_i}^5}{512 \pi^3 \Gamma_{\ell_i}} \, \frac{\left| M_{ijk} \right|^2}{\Lambda^4} \, ,
\end{equation}
where $\kappa = 2/3$ when there are two identical leptons in the final
state, and $\kappa = 1/3$ otherwise, and $m_{\ell_i}$ and
$\Gamma_{\ell_i}$ are the mass and decay width of the $\ell_i$ lepton,
respectively. The coefficient $M_{ijk}$ takes the form $C^L_{ij}
C^L_{kk}$ in case (i), 
$C^L_{ik} C^L_{jk}$ in case (ii).

One can now use the experimental limits on these LFV branching ratios
to derive bounds on $\Lambda$. Processes involving $C^L_{ee}$ are
strongly suppressed and thus they do not provide meaningful
bounds. This is the case of $\mu^- \to e^- e^- e^+$, $\tau^- \to e^-
e^- e^+$ and $\tau^- \to \mu^- e^- e^+$. In contrast, the {combined
  LHCb+BaBar+Belle} limit $\text{BR}(\tau^- \to \mu^- \mu^- \mu^+) <
1.2\times 10^{-8}$~\cite{Amhis:2014hma} translates into $\Lambda
\gtrsim 6.7$ TeV (in both scenarios, A and B). The other $\tau$ decay
modes lead to slightly less stringent bounds. Future B factories are
expected to improve on the search for $\tau^- \to \mu^- \mu^- \mu^+$,
with sensitivies to branching ratios as low as $\sim
10^{-9}$~\cite{Aushev:2010bq}, allowing us to probe NP scales up to
$\Lambda \sim 12$ TeV.

\section*{Conclusions and discussion}

In summary, we have suggested that the universality and flavor
violating $b \to s$ anomalies may be related to the pattern of
neutrino oscillations.
By assuming that the unitary transformation between weak and mass
charged lepton eigenstates is given by the leptonic mixing matrix
measured in neutrino oscillations we predict several lepton flavor
violating B meson decay rates.
This way we are led to the thrilling possibility that some of the rare
LFV B decay branching ratios correlate with the leptonic CP phase
$\delta$ that characterizes neutrino oscillations. Other lepton flavor
violating processes processes such as $\ell_i \to \ell_j \ell_k
\ell_k$ have been considered in a similar manner.
Improved measurements at Belle should probe new physics scale 
at the level $\Lambda \sim 12$ TeV. 
Relevant scenarios involve additional neutral currents, such as
schemes containing an extra $Z^\prime$ boson with lepton universality
violation in $B$
decays~\cite{Altmannshofer:2014cfa,Crivellin:2015mga,Crivellin:2015lwa,Sierra:2015fma},
or possibly some realizations of the electroweak symmetry ${SU(3)_C
  \otimes SU(3)_L \otimes
  U(1)_X}$~\cite{Singer:1980sw,Buras:2012dp,Boucenna:2014ela,Boucenna:2014dia,Buras:2014yna,Boucenna:2015zwa}. Such
schemes should be taken seriously should the observed hints in the B
sector persist.
Finally, we note that in this paper we have focussed on the case where
the violation of lepton universality is caused by NP in $B \to K \mu
\mu$, with negligible contributions to $B \to K e e$. The alternative
hypothesis is also plausible, though it has a lower constraining power
since the electron channel is experimentally somewhat less constrained
than the muonic one.

\section*{Note added}
A few days ago, an update of \cite{Altmannshofer:2014rta} was
presented in \cite{Altmannshofer:2015sma}.  While this would change
slightly the value of $\rho_{NP}$ used in our analysis, our main point
remains and the numerical results are also left essentially unchanged.

\section*{Acknowledgments}

We are grateful to Javier Virto, Jorge Martin Camalich, {and Marcin
  Chrz\k{a}szcz} for enlightening discussions. Work supported by the
Spanish grants FPA2014-58183-P, Multidark CSD2009-00064 (MINECO), and
the grant PROMETEOII/2014/084 from Generalitat Valenciana. SMB
acknowledges financial support from the research grant ``Theoretical
Astroparticle Physics'' number 2012CPPYP7 under the program PRIN 2012
funded by the Italian ``Ministero dell'Istruzione, Universit\'a e
della Ricerca'' (MIUR) and from the INFN ``Iniziativa Specifica''
Theoretical Astroparticle Physics (TAsP-LNF). AV acknowledges partial
support from the EXPL/FIS-NUC/0460/2013 project financed by the
Portuguese FCT.

\bibliographystyle{utphys}

\providecommand{\href}[2]{#2}\begingroup\raggedright\endgroup

\end{document}